Johnjoe McFadden and Jim Al-Khalili. School of Biomedical and Life Sciences, and Department of Physics, University of Surrey, Guildford, Surrey, GU2 5XH.

RE: Book Review of 'Quantum Evolution (Johnjoe McFadden) by Mathew J. Donald. Posted 4 Jan 2001, quant-ph/0101019.

This review critically assesses the popular science book 'Quantum Evolution' by Johnjoe McFadden (McFadden, 2001) and also the paper by McFadden and Al-Khalili (McFadden and Al-Khalili, 1999). The basic proposal pursued in both texts, which is criticised by Donald, is the possibility that quantum measurement may influence biological systems by dragging them along a particular dynamic trajectory. The proposal has two components:

1. That the dynamics of microbial biological systems may remain quantum coherent for biologically significant lengths of time (in the order of seconds).
2. That the interaction between a microbial genome and the cell's environment may constitute a quantum measurement that will perturb the dynamics of the system relative to the unmeasured system and thereby generate adaptive mutations.

We will examine the second proposal first, as it is more tractable to analysis.

As Donald describes, our proposal was made considering a simple model of mutations initiated by proton tunnelling from one site of DNA to another (tautomeric) site. The tautomeric proton has different base-pairing properties that may promote the insertion incorrect base during DNA replication to generate a mutation. We argued that for starved bacterial cells the proton tunnelling would be reversible; but in the presence of a substrate (lactose) that enables the cells to grow only if the DNA is in the mutant state, quantum measurement of the system by its environment (with lactose) will enhance the observed mutation rate.

Since writing the Biosystems paper and publishing 'Quantum Evolution' (and having to defend our ideas to our colleagues), we have reformulated the proposal in a setting that is more familiar to physicists. This is no way changes the arguments presented in our Biosystems paper, but we believe, makes the case clearer.

Consider a large numbers of particles – say 1 million – in a box containing a double potential well, Well X/Y. The particles all start off in Well X but can tunnel through an energy barrier into Well Y such that if you examine the wells at any time you will find nearly all the particles in Well X and an average of about 10 particles in well Y - essentially a stationary state solution with the probability density distribution constant in between measurements.

Now, you open the box to examine Well Y at any time and you will always find – more or less – the same numbers of particles: say about 10. The number of times you open the well and examine its contents will of course not change the distribution of particles between the wells. The X-Y transition is reversible (i.e. unlike radioactive decay).

However, if instead whenever you find particles in Well Y, you scoop them out and count them, and then close the box, you are making the X -> Y transition irreversible. The remaining particles within the wells will immediately re-equilibrate and more particles will tunnel from X to Y. If you do this scooping frequently, each scoop will retrieve about 10 particles. If you scoop often enough, you can shift all of the particles over to well Y. Essentially, not only do we make a measurement of the system but we remove particles from the shallow well, so that they are no longer part of the same quantum system after measurement.

To translate this back into the bacterial system we consider the box to be the bacterial genome and the two wells to represent the un-tunnelled and tunnelled state of the proton. So long as the proton tunnelling is irrelevant to the growth of the bacteria (no lactose present), then 'no-one' is looking in the well and the system will remain at the quantum level until thermal decoherence collapses the system into either state. After collapse due to thermal decoherence, the system will remain at the quantum level and the proton is again free to tunnel in either direction. Essentially, under these conditions, the tunnelling event is reversible and the proportion of cells with DNA in the mutant state will remain constant with time.

However, with lactose present, the tunnelling can cause a mutation that will allow the bacterium to grow (with lactose present). Under these conditions, the thermal field is not the only source of decoherence. Instead, the DNA is constantly 'being examined' for tunnelling events by the coupling of the mutant state with the presence of lactose. And CRUCIALLY, those examinations do not merely look, they 'capture' the tunnelling events by 'scooping out' those protons that promote growth and thereby collapse the bacterial DNA to the classical level to form a growing bacterial colony. The DNA of the remaining cells (the majority), in which the examination (decoherence) in the presence of lactose collapsed their protons into the unmutated state (still incapable of growth), remain at the quantum level and are free to tunnel once again into the mutant state to suffer successive rounds of decoherence.

The mutational process will thereby demonstrate decay kinetics with the proportion of cells in the mutant state increasing with time – but only if lactose is present: adaptive mutations. As we show in our paper, the enhancement in mutation rate caused by lactose will be proportional to the ratio of the two decoherence rates decoherence rates (+/- lactose).

The proposed system is in many ways analogous to a quantum ratchet in which an asymmetric potential can be used to induce directional particle flow through a tunnelling barrier (Linke et al., 1999). It also bears similarities to 'environmental engineering' of quantum states as discussed by Pablo Paz (Pablo Paz, 2001) and demonstrated, for example, by Wineland's group (Myatt et al., 2000). In these situations, decoherence is controlled by manipulation of the environment, to select certain states from a quantum superposition. We propose that the level of lactose in the growth media similarly engineers the level of environment for selected mutational states of the bacterial DNA.

In a sense, the system is also analogous to alpha decay. A Geiger counter may measure the decay but it does not change the rate of alpha decay since it does not 'capture' the tunnelled state. It is the escaped particle's entanglement with its environment that 'removes' the leaving particle and thereby makes the system irreversible and the decay progressive. If the decay did not involve a particle escaping from the nucleus, then the two states would rapidly reach an equilibrium depending of their energies, within the nucleus. The distribution between the decayed and undecayed state would be constant with time – the decay would not be progressive. Under these circumstances, removal of one of the decay product by some kind of irreversible coupling with the environment (as of course occurs in alpha decay) would accelerate the decay.

This model presented is plausible and perfectly compatible with known physics. However, as Donald points out, the model does depend on the warm wet biological system remaining coherent for lengthy periods of time (our second proposal). Donald does not believe this is possible. In both our paper and McFadden's book, this issue is not dodged. We fully understand the difficulties involved and pointed to a number of possible solutions. We cited experimental evidence for quantum coherence in living systems and pointed out that the difficulties in accounting for quantum coherence in, for instance, high temperature superconductors, indicating there are many gaps in our understanding of decoherence in complex systems. Recent demonstrations of 'dynamical tunnelling', as discussed by Eric J. Heller (Heller, 2001), similarly indicates that our understanding of coherence within dynamic systems is far from complete. Experimental systems that have been used to examine decoherence are very different from the highly structured environment of a living cell; and it is possible that estimates of decoherence rates gained from inanimate systems do not apply to living cells. However, both in the paper and book we took pains to indicate that this was only a hypothesis and remains to be tested. Recent experiments (Schon et al., 2001)demonstrating superconductivity in doped fullerene (C60) at 117K (with indications that higher temperatures may be attainable) indicates that certain organic structures may indeed promote quantum coherence. Transport of charges along the DNA double helix by hole transfer through quantum tunnelling has also been recently demonstrated (Giese et al., 2001), as has coherent proton tunnelling in a hydrogen bonded network (Horsewill et al., 2001). Coherent reaction dynamics have also recently been demonstrated for the cytochrome *c* oxidase enzyme (Liebl et al., 1999). Although none of these observations is proof of the existence of quantum coherence in living biological systems, they do offer tantalising indications that our speculations were not entirely unfounded.

Donald very specifically criticises our use of proton relaxation rates to insert into Zurek's equation that we use to estimate decoherence, which he describes as 'an error'. The value we needed was the relaxation time for a coding proton in DNA, a measure of the speed of energy dissipation. As we state in our paper, this value is completely unknown and we go on to state that 'some measure of the possible range of energy dissipation times for protons in living systems may be gained from examination of proton relaxation rates in biological materials, measured by NMR. … Although the exact relationship between the NMR $T_1$ value and the relaxation rate is far from clear, they are both a measure of the rate of energy exchange between a proton and its environment.' Donald may disagree with these statements but they do not constitute and 'error'. Indeed, measurement of $T_1$ values was recently utilised to examine the extent of coherent proton tunnelling in an artificial hydrogen bond network (Horsewill et al., 2001). We clearly state the difficulties with estimates of relaxation rates, but rather than pulling a value out of the air, we incorporated an experimentally derived value of a parameter that at least has some similarity to the one under consideration.

Donald rightly points out that our decoherence rate estimate was for only a single proton whereas a mutation will involve many more particles. Unfortunately there is currently no method available that would allow us to calculate decoherence rates for such a complex system from dynamic principles alone, so our argument was indeed weak at this point, merely arguing that 'DNA, RNA and protein will differ only at single residues and, therefore, only involve relatively small-scale atomic displacements for very small numbers of particles. We propose that under these conditions, quantum coherence persists within the cell …'. Essentially we were proposing that, despite the fact that the living cell is indeed a highly complex system that interacts with its environment, certain of its degrees of freedom (those consequent upon the proton tunnelling event) might remain sufficiently isolated to maintain quantum coherence.

The degree of isolation of those degrees of freedom will depend on the leakage of information - from the cell to its environment – that would betray the positions of the particles involved. Crucially, despite the fact that the cell clearly does exchange lots of information with its environment, much of the information is not capable of betraying particle position, because of the relatively long wavelengths involved. For instance, the photons emitted by living cells (biophotons) are characteristically in the visible or infrared spectrum and thereby inefficient sources of information of angstrom level particle displacements within DNA or proteins. The other main source of information leakage from a living cell would be phonons. We are currently attempting to gain some estimate of the amount of information carried by phonons across cell membranes, but there is unfortunately very little data available on the phonon spectrum of living tissue. Clearly this is an issue we must address. With so much uncertainty, although Donald may consider it highly unlikely that quantum coherence could be retained under these conditions, our argument cannot be said to be 'wrong'.

We have not here attempted to defend the analogy of the system with the inverse Zeno effect. In our paper we claim only that 'The phenomenon bears many similarities to the inverse quantum Zeno effect'. In McFadden's book 'Quantum Evolution', the analogy with the inverse Zeno effect is explored more fully and is indeed proposed to be involved in the origin of life, but no detailed model is given. Donald points out a many difficulties with a biological version of the inverse Zeno effect, but, as he rightly points out, there are many difficulties with all explanations of the origin of life. If the emergence of life depended on an unlikely sequence of measurements that invoked the inverse Zeno effect, then it may yet be the most plausible 'origin of life' scenario.

In summary, in his criticisms, Donald fails to understand the key mechanism of our hypothesis and overlooks many of the qualification we made in our analysis. However, we are grateful for Donald for his serious attempt to consider the dynamics of biological systems at the quantum level. As Donald points out, the credibility of our case hangs on the feasibility of finding quantum states existing in living cells. We would welcome any suggestions as to how to explore this issue further, either theoretically or experimentally.

**References**

Giese, B., Amaudrut, J., Kohler, A.K., Spormann, M., and Wessely, S. (2001). Direct observation of hole transfer through DNA by hopping between adenine bases and by tunnelling. *Nature* **412,** 318-320.

Heller, E.J. (2001). Air juggling and other tricks. *Nature* **412,** 33-34.

Horsewill, A.J., Jones, N.H., and Caciuffo, R. (2001). Evidence for coherent proton tunneling in a hydrogen bond network. *Science* **291,** 100-103.

Liebl, U., Lipowski, G., Negrerie, M., Lambry, J.C., Martin, J.L., and Vos, M.H. (1999). Coherent reaction dynamics in a bacterial cytochrome c oxidase. *Nature* **401,** 181-184.

Linke, H., Humphrey, T.E., Lofgren, A., Sushkov, A.O., Newbury, R., Taylor, R.P., and Omling, P. (1999). Experimental tunnelling ratchets. *Science* **286,** 2314-2317.

McFadden, J. (2001). Quantum Evolution (New York: WW Norton).

McFadden, J. and Al-Khalili, J. (1999). A quantum mechanical model of adaptive mutation. *Biosystems* **50,** 203-211.

Myatt, C.J., King, B.E., Turchette, Q.A., Sackett, C.A., Kielpinski, D., Itano, W.N., and Monroe, C.W.D.J. (2000). Decoherence of quantum superpositions through coupling to engineered reservoirs. *Nature* **403,** 269-273.

Pablo Paz, J. (2001). Protecting the quantum world. *Nature* **412,** 869-870.

Schon, J.H., Kloc, Ch., and Batlogg, B. (2001). High-temperature superconductivity in lattice-expanded $C_{60}$. *Science* **293,** 2432-2434.